\documentstyle[preprint,prl,aps]{revtex}

\begin{document}
\draft
\preprint{}
\title{ Estimates of electronic interaction parameters for La$M$O$_3$
compounds ($M$=Ti-Ni) from {\it ab-initio} approaches}
\author{Priya Mahadevan$^{1,2}$, N. Shanthi$^{1}$ and D. D. Sarma$^{1,*}$}
\address{$^1$ Solid State and Structural Chemistry Unit, Indian Institute of 
Science, Bangalore 560012, India}
\address{$^2$ Department of Physics, Indian Institute of 
Science, Bangalore 560012, India}
\date{\today}
\maketitle
\begin{abstract}

We have analyzed the {\it ab-initio} local density approximation
band structure calculations for the family of perovskite
oxides, La$M$O$_3$ with $M$=Ti-Ni within a parametrized nearest
neighbor tight-binding model and extracted various interaction
strengths. We study the systematics in these interaction
parameters across the transition metal series and discuss the
relevance of these in a many-body description of these oxides.
The results obtained here compare well with estimates of these
parameters obtained via analysis of electron spectroscopic
results in conjunction with the Anderson impurity model. The dependence
of the hopping interaction strength, $t$, is found to be approximately
$r^{-3}$.
\end{abstract}

\pagebreak

\noindent {\bf Introduction}

Electronic structures of transition metal oxides have
attracted a great deal of attention in recent years,
arising from their unusual electronic and magnetic
properties. Simultaneous presence of strong electron-electron
interaction within the transition metal 3$d$ manifold
and a sizeable hopping interaction strength between
the 3$d$ and oxygen 2$p$ states is primarily responsible
for the wide range of properties exhibited by these
compounds. Often the presence of a strong intraatomic
coulomb interaction strength makes a single-particle
description of such systems inadequate, necessitating
a model many-body Hamiltonian approach \cite{ZSA,Sarma}.
However, such parametrized approaches require a prior
knowledge of the various electronic interaction
strengths, such as the intraatomic Coulomb interaction strength,
$U_{dd}$, the charge-transfer energy, $\Delta$, and the
metal $d$-oxygen $p$ hopping interaction strength, $t$,
in order to provide a description of the ground
state electronic structure.  Traditionally, various
high-energy spectroscopic results in conjuction
with model Hamiltonian calculations have provided
estimates for such interaction strengths \cite{core1,core2,core3,core4}.
However, it is well-known that wide ranges of parameter
values are compatible with experimentally obtained spectra,
leading to non-unique solutions \cite{Unpub}. This arises
from the fact that the effect of a change in one parameter
on the calculated spectrum can often be compensated by
suitably changing another parameter, such that the
the various solutions  are indistinguishable
within experimental uncertainties. This problem has given
rise to different sets of estimates of interaction
parameters from different groups for the same compound.
In this context, it is highly desirable to obtain
independent estimates of various interaction strengths
from methods other than those using spectroscopic
results. Even if only one interaction strength can be
reliably estimated from any other method, the problem
of non-uniqueness in analysing experimental spectroscopic
result is considerably eliminated and such an approach
will yield a more consistent description of the electronic
structure.
 
Recently we have shown \cite{PRL} that the electronic and magnetic
structures of a particular class of transition metal oxides,
namely the 3$d$ transition metal perovskite oxides of the
general formula La$M$O$_3$ ($M$=Cr-Ni) are described
very accurately within {\it ab-initio} approaches
based on local density approximation (LDA). Futhermore,
it is also seen that the various excitation spectra,
such as x-ray photoemission (XP) spectra of the
valence band region and bremsstrahlung isochromat
(BI) spectra of these compounds are also described
well within the band structure approach. This success
suggests that such band structure results could be
useful in providing reliable estimates of various
interaction strengths in this interesting class
of compounds which have attracted a lot of attention
in recent times \cite{Sys}. 
Estimates of various interaction strengths
based on {\it ab-initio} calculations have indeed been
carried out in the past, for example
for the 3$d$ transition metal monoxide series
\cite{Mattheiss}, La$_2$NiO$_4$ \cite{ref1},
La$_2$CuO$_4$ \cite{ref2}, the
4$d$ transition metal oxide, Sr$_2$RuO$_4$ \cite{Oguchi}
and Mn doped CdTe \cite{CdTe}.
We report here the results of our analysis
for the entire series of compounds
belonging to the La$M$O$_3$ family with $M$=Ti-Ni.
Wherever possible, we compare the results obtained
from the present approach with those obtained by 
analysis of various spectroscopic results
already existent.

\noindent {\bf Methodology}

We have performed detailed band structure calculations
within the linearised muffin-tin orbital (LMTO)  method
using the atomic sphere approximation (ASA) \cite{LMTO}.
Spin restricted calculations were performed for the real
crystal structures experimentally observed. Thus, no
attempt has been made to determine the lattice constants
from total energy calculations. This is justified in 
the present context, since our primary aim here is to obtain
strengths of the electronic interactions in the real systems
with the observed lattice constants. In the case of LaCoO$_3$,
however, we have performed calculations for various values
of the lattice parameter in order to deduce the dependence of
the hopping interaction strengths on distances. Analysis of
these results shows that the present method predicts the lattice
parameters with nearly 1\% accuracy. 
LaTiO$_3$ \cite{Ti}, LaVO$_3$ \cite{V}, LaCrO$_3$
\cite{Cr} and LaFeO$_3$ \cite{Fe} in the observed Pbnm structure
were calculated with twenty atoms in the unit cell,
while LaCoO$_3$ \cite{Co} and LaNiO$_3$ \cite{Ni}
calculations were performed with ten atoms in the unit cell
in the real R$\bar{3}$c structure. In the case of LaMnO$_3$,
the real  crystal structure is Pnma stabilised by a 
Jahn-Teller distortion around the Mn-ions \cite{Mn1}.
In contrast to all other La$M$O$_3$ compounds,
this leads to a distorted octahedra of oxygens with two distinct
Mn-O distances. Since, this would introduce additional
parameters in the analysis, we have preferred to
calculate the band structure of LaMnO$_3$ with an earlier
reported R$\bar{3}$c idealised structure, where the Jahn-Teller
distortion is suppressed \cite{Mn2}. Such approximations
have been made in the past \cite{Fujimori} and we do not
expect to introduce any significant error in the
parameters obtained, since the change in the interaction
parameters is expected to be small for the small
distortions observed.

Various interaction strengths can be estimated by mapping
the band dispersions obtained from a parametrized tight-binding
model onto those from {\it ab-initio}
band calculations \cite{Mattheiss}. Since, we are primarily
interested in the states arising from transition metal
$d$-oxygen $p$ interaction which dominate the
occupied density of states as well as parts of the unoccupied states
near the Fermi energy, E$_F$, the tight-binding Hamiltonian
consists of the bare energies of the transition metal $d$
(${\epsilon}_d$) and oxygen $p$ (${\epsilon}_p$) states and hopping
interactions between all these states.  The nearest 
neighbor hopping interactions were
expressed in terms of four Slater-Koster parameters,
namely $pp\sigma$, $pp\pi$, $pd\sigma$ and $pd\pi$ \cite{Slater}.
It is further necessary \cite{Matth1} to include an interaction,
$sd\sigma$, between the transition metal $d$ and oxygen
$s$ states in order to simulate the splitting between the
$t_{2g}$ and $e_{g}$ states at the ${\it \Gamma}$ point,
 since the degeneracy  of these two states
is not lifted by the $p-d$ interaction at this symmetry
point. In general, it is also necessary
to include the effect of non-orthogonality of atomic
functions located at different sites by considering
the overlap matrix in such tight-binding approaches\cite{Matth1}.
 On the other hand,
the tight-binding part of any parametrised many-body
Hamiltonian, such as the Hubbard Hamiltonian,
ignores the overlap matrix with an assumption
of orthogonality of the basis functions.
Since we are eventually interested in obtaining estimates
of interaction strengths that enter such many-body 
Hamiltonians as parameters, we have carried out the
fitting of the LMTO dispersions within two separate
tight-binding models, one with and the other
without the assumption of orthonormal basis functions.

Since we take into account only the transition metal
$d$ and oxygen $p$ states, there are 56 bands to be
fitted in the Pbnm structure and 28 bands in the
R$\bar{3}$c structure. LMTO-ASA results indicate that
several of the low-lying bands within the nominal
oxygen $p$-bandwidth have non-negligible
contributions from La derived states. Thus, we have
not taken into account some of these bands in the 
fitting procedure. We included the top 38 bands in the case
of the Pbnm structure and 19 in the R$\bar{3}$c structure.
In the case of LaTiO$_3$, the top 8 bands with mainly Ti
$d$ character overlap extensively with other bands 
having primarily La character. Thus we could not include these
bands in the fitting procedure.
We also checked the effect of including only the primarily
transition metal $d$-derived bands, (top 20 in Pbnm and
10 in R$\bar{3}$c structures), leaving out all the primarily
oxygen $p$-derived bands, from the fitting procedures. 
However, in the case of LaTiO$_3$ we have not carried out
such an analysis due to the extensive overlap of Ti $d$ and La
derived bands mentioned above.
Results of these different procedures, namely assuming the orthogonality
or non-orthogonality of the atomic basis functions
and employing different limited sets of bands, are consistent
with each other.

\noindent {\bf Results and discussion}

Total density of states as well as partial densities of
Mn $d$ and O $p$ states in LaMnO$_3$ states obtained
from LMTO-ASA calculations are shown in  Fig. 1. We
show the corresponding band dispersions  along various
symmetry directions in Fig. 2a. The
zero of the energy scale corresponds to the Fermi energy,
E$_F$. A set of  two bands can be seen close to 3 eV at the
${\it \Gamma}$ point and dispersing to higher energies along both
${\it {\Gamma}-L}$ and ${\it {\Gamma}-Z}$ directions, becoming larger 
than 4 eV halfway to the zone boundaries. These two
bands are primarily La derived with negligible Mn $d$
contributions (see Fig. 1) and we do not discuss these
any further here. Next, one can observe a group of
four strongly dispersing bands in Fig. 2a between
about 0.5 and 3.5 eV giving rise to the approximately
rectangular DOS marked A in Fig. 1. These bands are derived primarily
from Mn $d$($e_g$)-oxygen $p$-admixture with dominant Mn $d$
character. Below the $e_g$ bands, there is a group of
6 bands spread between -1 and 0.5 eV with
considerably less dispersion than the $e_g$ bands (Fig. 2a)
giving rise to a narrow DOS marked B in Fig. 1.
Various partial densities of states in this energy range
suggest that this group has
primarily Mn $d$ character with some finite oxygen $p$
admixture (Fig. 1). From these observations it is clear
that this group of bands arises mainly from Mn $d$ ($t_{2g}$)
states. Both $e_g$ and $t_{2g}$ bands discussed so far arise
from Mn $d$-oxygen $p$ antibonding combination, with
dominant Mn $d$ character. Thus, these bands are termed
antibonding $e_g^{*}$ and $t_{2g}^{*}$ bands
respectively. At still lower energies, we find an
energy region with DOS between about -2.7
and -4.5 eV which is almost completely contributed
by oxygen $p$-character with very small Mn $d$
character. These features in the DOS are marked as
C in Fig. 1 and arise from essentially oxygen-oxygen
interactions with a non-bonding character with
respect to Mn $d$- O $p$ interactions. The corresponding
dispersions of the eight bands can be observed in
Fig. 2a. Dispersions of the remaining ten bands can be seen
between -4 and -7.5 eV. The corresponding density of
states (marked D) peaks at about -5.5 eV and has primarily
oxygen $p$-character (see Fig. 1). However, finite Mn $d$
admixture is also observable in this energy range.
Thus, these states arise from oxygen $p$-Mn $d$ bonding
interactions and are the bonding  counterparts of the Mn
$d$-dominated antibonding $e_g^*$ and t$_{2g}^*$ bands
at higher energies. These interpretations are consistent with
the description of bonding in La$M$O$_3$ compounds which have
been discussed in detail in recent times \cite{Bond1,Bond2,Bond3}. 

In order to map the band dispersions obtained from
LMTO-ASA calculations onto the tight-binding model,
we have used all the ten primarily Mn $d$-derived
$e_g^*$ and $t_{2g}^*$ bands as well as the top nine
oxygen $p$-bands, as explained in the previous section. 
The best fit tight-binding dispersions are shown in Fig. 2b
for comparison with the {\it ab-initio} calculated
dispersions in Fig. 2a. It is evident from Fig. 2
that good agreement is obtained between
the LMTO and tight-binding results; this is
particularly true of the bands related to the
$t_{2g}^*$ and $e_g^*$ distributed between -1 and 3.5 eV.
This is significant, since the lower bands related
primarily to oxygen $p$-states are completely
filled in all these oxides and the electronic
and magnetic  properties in these cases are
controlled entirely by the $t_{2g}^*$ and $e_g^*$
bands. Moreover, the present result 
suggests that the electronic structure of these
transition metal oxides can be well described 
in terms of models involving only the transition
metal $d$ and oxygen $p$ states (i.e $d-p$ models), as is the
usual procedure.

While the above mentioned case of LaMnO$_3$ was calculated
within the R$\bar{3}$c structure, we show an example
of LMTO-ASA DOS and band dispersions for the Pbnm structure
in Figs. 3 and 4a respectively for LaFeO$_3$.
Once again, four groups of DOS features are easily
identified in Fig. 3 for LaFeO$_3$. These are marked A through
D and have the same origin as discussed in the case of
LaMnO$_3$ shown in Fig. 1. In the case of the Pbnm structure
of LaFeO$_3$ however, there are twenty atoms in the unit cell,
thereby doubling the number of bands in comparison
with the previously discussed R$\bar{3}$c structure. Thus, within 
the $d-p$ part of the electronic structure alone, there are  fifty-six
bands; eight related to the $e_g^*$ bands, twelve to the
$t_{2g}^*$ bands, sixteen related to the so-called
non-bonding oxygen $p$-parts and twenty related to the
bonding states  of the Fe $d$-O $p$ interactions.
For the sake of clarity, we show the  dispersions of those bands
which are related to the $e_g^*$ and $t_{2g}^*$
bands  with primarily Fe $d$ character
along various symmetry directions in Fig. 4a.
The strongly dispersing group of eight bands
(more easily recognisable along the ${\it {\Gamma}-R}$ direction)
between about 0.5 eV and 3 eV arise from the $e_g^*$
bands, while the considerably more flat twelve bands
between -1 and 0.5 eV comprise the $t_{2g}^*$ region.
These band dispersions alongwith the sixteen non-bonding
oxygen $p$ bands were fitted with $d-p$ only
tight-binding model along all symmetry directions
in the same way as in the case of LaMnO$_3$.
The resulting best fit tight-binding results are
shown in Fig. 4b. Comparison of Figs. 4a and b
indicate that the $d-p$ only nearest neighbor
tight-binding model describes the band dispersions
observed within the {\it ab-initio} calculation
quite well.

The parameters for the best fits 
to the LMTO band dispersions for all the La$M$O$_3$
compounds with $M$=Ti-Ni are given in Table I.
While the main entries in this Table are for the cases where
$d$ and $p$ bands were fitted, the numbers in parentheses
are obtained by fitting only the $e_g^*$ and $t_{2g}^*$
bands. These two sets of estimates are very similar 
exhibiting some systematic changes across the transition 
metal series. We have plotted the
strengths of various hopping interactions across the 3$d$
transition metal series in Fig. 5. It is reasonable to
expect that the variations in the hopping interactionwill be related to the
changes in the relevant atomic distances. Thus, we have also
shown the experimentally observed oxygen-oxygen ($r_{O-O}$) and the 
metal-oxygen ($r_{M-O}$) distances in the inset of Fig. 5. Both $r_{O-O}$
and $r_{M-O}$ appear to be larger in LaFeO$_3$ compared to the overall 
trend. This may arise from the stability of the half-filled 
high-spin $d^5$ ionic configuration of Fe$^{3+}$ in this compound. The 
decrease of these distances for LaCoO$_3$, just after LaFeO$_3$,
is clearly related to the low-spin configuration of the Co$^{3+}$ ion. 
Fig. 5 shows that the magnitude of $pp\sigma$ does not show any significant 
variations between V and Mn, then decreases for Fe followed
by a substantial increase for Co. These variations
in $pp\sigma$ strength can be easily related to the changes in the nearest
neighbour oxygen-oxygen distances (r$_{O-O}$) in the La$M$O$_3$
compounds; r$_{O-O}$ substantially increases for LaFeO$_3$ giving 
rise to the observed decrease in $pp\sigma$ in this compound. 
For LaCoO$_3$, r$_{O-O}$ is seen to be the smallest
in the La$M$O$_3$ series and for LaTiO$_3$, $r_{O-O}$ is the
largest; consequently the corresponding 
strengths of $pp\sigma$ are the largest and
the smallest respectively in the series. Various
transition metal $d$-oxygen orbital interactions,
namely $sd\sigma,~pd\sigma$ and $pd\pi$ also exhibit
systematic variations across the series, the
qualitative behaviour of these three interactions
being quite similar to each other (Fig. 5).
These interactions, besides being influenced
by the transition metal-oxygen distance
(r$_{M-O}$) shown in the inset, are also
influenced by the spatial extent of the transition metal
$d$ orbitals. It is well-known that the $d$-orbitals
contract across the transition metal series. Thus,
the decreasing trends in the hopping interaction
strengths between the transition metal $d$ and
oxygen orbitals between Ti and Mn, inspite of a small
decrease in r$_{M-O}$ arise from the $d$-orbital
contraction across the series. Interestingly, the
interaction strengths are larger for LaCoO$_3$
than LaFeO$_3$. This is clearly related to the
pronounced decrease of r$_{M-O}$ between LaFeO$_3$
and LaCoO$_3$, associated with the low-spin configuration
in LaCoO$_3$ in contrast to the high-spin configuration
in all the other La$M$O$_3$ compounds with $M$=Ti-Fe.

Besides the various hopping interaction strengths, the
bare energy difference, 
($\epsilon_d-\epsilon_p$),
between the transition metal $d$ and oxygen $p$
orbitals exhibits a monotonic decrease across the
La$M$O$_3$ series, as shown in Fig. 6. This is indeed
the expected trend, since the transition metal $d$
level is increasingly stablized with respect to 
the oxygen $p$-orbital due to the increasing nuclear
potential with increasing atomic number of the transition
metal ion. 

The results summarized in Table I and Figs. 5 and 6
are obtained by fitting the LMTO band dispersions to the results
of the tight-binding model with finite overlap between the orbitals
at different atomic sites. However, most of the model many-body
Hamiltonian approaches assume an orthogonal basis set;
thus, the results in Table I cannot be directly
used to estimate the parameter strengths appearing in such models.
In order to provide estimates for such interaction strengths
that parametrize the many-body calculations, we have also
fitted the LMTO band dispersions with the results of a tight-binding
Hamiltonian assuming an orthogonal basis.
Thus, the tight-binding model corresponds
to the one-electron part of the multiband Hubbard model
involving all the transition metal $d$ and oxygen $p$
orbitals, suitable for the La$M$O$_3$ series. The
resulting estimates of the various interaction
strengths corresponding to the best simulation of LMTO
band dispersions are given in Table II. 
A comparison of Tables I and II show that
the estimates of various parameters are quite similar
in the two cases, justifying the assumption of an
orthogonal basis in describing the electronic structures
of these compounds. This is further supported by the
fact that the orbital overlaps required to optimize
the simulation of LMTO band dispersions are generally
very small. We also find that the parameter
values summarized in Table II exhibit similar systematic
trends across the transition metal series as those in Table I
(see Figs. 5 and 6).

In order to verify the relevance of the interaction parameters
thus estimated, we note that there have been several
attempts in the past to estimate many of these
from various high-energy spectroscopic results
in conjunction with different many-body approaches \cite{Sarmarev}.
It is well known that such estimates are often non-unique
\cite{Unpub}; however it is believed that in particular,
various hopping interaction strengths can be estimated
with a fair degree of accuracy from such approaches. Thus,
we compare the estimates of the transition metal $d$-
oxygen $p$ interaction, $pd\sigma$ obtained from high-energy
spectroscopy and the present approach. 
Two different groups have systematically
worked on obtaining these experimental estimates for La$M$O$_3$ 
compounds. In Fig. 7, we have
plotted the experimentally obtained estimates of
$pd\sigma$ from the Tokyo group \cite{core1} and the Bangalore
group \cite{core3} against the 
estimates obtained in the present work. 
We show the results from these two groups using different
symbols; we have drawn two broken lines as guides to the eye
for the average overall variation of $pd\sigma$ across the series, as
obtained by each of the two groups. The solid line in Fig. 7 with a
unity slope and no intercept, represents the behaviour
of $pd\sigma$, if the experimental estimates and the present
estimates were identical in every case. Obviously, the
experimental estimates are somewhat different from the
calculated ones. However, it is clear that the average
variations of $pd\sigma$ across the series obtained by these
groups are very similar to that suggested by the present
calculations; this is indicated by the fact that the broken
lines representing the average variation of experimentally
obtained $pd\sigma$ are approximately parallel to the solid
line within the accuracy of experimental estimates. 
Moreover, while the estimate of $pd\sigma$ from one
group is higher than the calculated ones, the estimate from the 
other is consistently lower. This arises from the non-uniqueness
of the parameter strengths estimated from experimental results
discussed earlier. More importantly, it is 
obvious that $pd\sigma$ values estimated in the
present work are also compatible with the high-energy 
spectroscopic data, since values both larger and
smaller than the calculated values provide
satisfactory description of the experimental
observations. Thus, it is desirable to constrain
the strength of $pd\sigma$ to the values calculated here
on the basis of {\it ab-initio} approaches; this will
avoid the problem of non-uniqueness normally
encountered in such analysis \cite{Unpub} and will help
in obtaining considerably better defined estimates
of other parameters in the model.

The estimates of ${\epsilon}_d-{\epsilon}_p$ obtained by
analysing the LMTO results can be related 
in an approximate way to the charge
transfer energy, $\Delta$ defined \cite{ZSA} within
the Anderson impurity Hamiltonian which is often used
to provide a many-body description of the
electronic structure of these oxides. For this, we first note that the
eigen values,  ${\epsilon}_d$ and ${\epsilon}_p$
in an LDA calculation are not directly related
to the orbital ionization energies. In order to see
the relationship between the charge transfer 
energy and the eigen values, we note that the total
energy $E$ within the LDA calculation can be expressed
as a Taylor series expansion in terms of the electron occupancies,
$n_i$, of the various levels $i$:
$$E(...n_i...) = E_0 +\sum_{i} b_in_i +\sum_{ij} a_{ij}
n_i n_j + ... $$ 
Retaining terms upto the second order in the above expansion,
it is easy to show \cite{Sarmarev} that the coefficients
$a_{ij}$'s are related  to various electron-electron
interactions. The atomic orbital eigen value {$\epsilon_i$} of the level
$i$ is related \cite{SlaterXA} to the total energy as
$$ {\epsilon}_i = {{\partial E}\over{\partial n_i}}=2[\sum_{j}
a_{ij}n_j]+b_i $$ 
Then the charge transfer energy, {$\Delta$}, defined as
$${\Delta}= E(...n_d+1...n_p-1...)-E(...n_d...n_p...)$$
is readily given by 
\begin{eqnarray}
{\Delta} = {\epsilon}_d-{\epsilon}_p + {1\over{2}}U_{dd}
\end{eqnarray}
Thus, ${\Delta}$ and the difference in the orbital eigen values,
${\epsilon}_d-{\epsilon}_p$ are linearly related with the difference
being half of the intraatomic Coulomb strength within the 3$d$
electrons. In Fig. 8, we have plotted the values of $\Delta$
obtained from various high energy spectroscopic results
as a function of the orbital energy difference, ${\epsilon}_d-{\epsilon}_p$,
calculated in the present work. The straight line with a unity slope
represents the expected dependence of $\Delta$ on ${\epsilon}_d-{\epsilon}_p$
in absence of Coulomb interaction. With the sole exception
of the case of V, all experimental estimates of $\Delta$
is larger than that represented by the straight line; this
clearly indicates the presence of finite Coulomb
interaction strengths in all these cases. We however
would not like to estimate $U_{dd}$ from the difference
between the experimentally estimated $\Delta$ values
and the value (represented by the  straight line in Fig. 8)
expected in absence of $U_{dd}$, since the uncertainty in
estimating $\Delta$ from experiment can be very large
(as large as $\pm$ 2 eV). This is also possibly
the reason why the $\Delta$ value in the case of the vanadium
compound has been estimated lower than the straight line. 
The present results suggest that the estimate
of ${\epsilon}_d-{\epsilon}_p$ in this work can be used to
guide the choice for $\Delta$ via the equation 1.
Essentially, equation 1 gives a relationship between
$\Delta$ and $U_{dd}$, since ${\epsilon}_d-{\epsilon}_p$
is estimated here, this leaves only one parameter
($U_{dd}$ or $\Delta$) to be determined from experimental 
results.

On many occassions, it is desirable to know how $pd\sigma$
changes with the distance between the transition metal
and the oxygen atom in a compound. This may be very useful
in describing the electronic structure of any compound
under pressure, or in the case of compounds, such as
LaMnO$_3$, where various transition metal-oxygen
distances are observed. Thus, there have been many
attempts to express the functional dependence between
$pd\sigma$ and $r_{M-O}$. It was shown earlier that 
the interatomic matrix-elements are supposed to scale
with distance as 1/$r^{l+l'+1}$ \cite{scale}, where l and l$'$ are the
angular momenta of the orbitals involved. Thus, for $p-d$
interactions, the matrix elements scale as 1/$r^4$. In order to obtain an
{\it ab-initio} estimate of the functional dependence,
we have calculated the band structure of LaCoO$_3$ with
different lattice parameters. The different sets of
band dispersions have been analysed in terms of the same 
tight-binding Hamiltonian with overlap included.
The parameters in each case were obtained by the 
previously discussed least-squared error
fitting procedure. We have plotted the $pd\sigma$
thus obtained as a function of $r_{Co-O}$ for the
eight different calculations in a log$_{10}$-log$_{10}$ plot. The
smallest $r_{Co-O}$ corresponds to a $4\%$
contraction while the largest value corresponds to
a $9.9\%$ expansion over the equilibrium $r_{Co-O}$.
Over this entire range, log$_{10}$($pd\sigma$) is found to vary
approximately linearly with log$_{10}$($r_{Co-O}$) with a slope of
-3.04$\pm$0.02. This implies that $pd\sigma$ varies
approximately as 1/$r_{Co-O}^3$ which in other terms is 
1/$r^{l+l'}$ in contrast to the earlier expectation
of 1/$r^{l+l'+1}$ dependence. 

In conclusion, the electronic structure of the La$M$O$_3$ series have been
studied by mapping the results of a tight-binding 
Hamiltonian with nearest-neighbor interactions onto the {\it ab-initio}
band structure results for $M$=Ti-Ni.
All the interaction parameters are found to exhibit systematic
changes across the transition metal series. The estimates
of various interaction strengths compare well with estimates
obtained from spectroscopic data, wherever available. The hopping
interaction strength, $pd\sigma$ is found to depend on the
transition metal-oxygen separation, $r$ approximately as $r^{-3}$.

\noindent {\bf Acknowledgment}

D.D.S. thanks Dr. M. Methfessel, 
Dr. A. T. Paxton, and Dr. M. van Schiljgaarde for making
the LMTO-ASA band structure program available and Dr. S. Krishnamurthy
for initial help in setting up the LMTO-ASA program. 
N.~S. and P.~M. thank the Council of Scientific and Industrial Research, 
Government of India for Research Fellowships.

\pagebreak

\section{figure captions}

Fig.1 The LMTO DOS for LaMnO$_3$ (solid line) alongwith
the partial DOS for the Oxygen $p$ states (dotted line)
as well as for the Mn $d$ states (dashed line).

Fig.2 (a). The LMTO band dispersions, and (b) the best-fit TB band
dispersions for LaMnO$_3$ along the 
symmetry directions ${\it \Gamma}L$, $LF$, $FZ$, $Z{\it \Gamma}$.

Fig.3 The LMTO DOS for LaFeO$_3$ (solid line) alongwith the
partial DOS for the Oxygen $p$ states (dotted line) as well as
for the Fe $d$ states (dashed line).

Fig.4 (a). The LMTO band dispersions, and (b) the best-fit TB band
dispersions for LaFeO$_3$ along the 
symmetry directions ${\it \Gamma}X$, ${\it \Gamma}Y$, ${\it \Gamma}Z$ 
and ${\it \Gamma}R$.

Fig.5 The variations of the interaction strengths $sd\sigma$,
$pp\sigma$, $pp\pi$, $pd\sigma$ and $pd\pi$ across the transition
metals series, Ti-Ni. The inset shows the interatomic distances
between the metal-oxygen ($r_{M-O}$) and the oxygen-oxygen 
($r_{O-O}$) sites in  \AA for the series.

Fig.6 The variation of ${\epsilon_p}$-${\epsilon_d}$ across
the transition metals series, Ti-Ni.

Fig.7 A comparison of the calculated $(pd{\sigma})_{calc}$ to
the values estimated from experiments, $(pd{\sigma})_{expt}$ 
by different groups (Tokyo and Bangalore). The solid line has been drawn for 
$(pd{\sigma})_{calc}$ equal to $(pd{\sigma})_{expt}$ and the dashed lines
indicate the trends seen in the data from different groups.

Fig.8 A comparison of the calculated value of ${\epsilon}_d-{\epsilon}_p$
with the experimentally obtained ${\Delta}$. The 
solid line represents ${\Delta}$=${\epsilon}_d-{\epsilon}_p$.

Fig.9 The variation of log$_{10}(pd\sigma)$ with
log$_{10}(r_{Co-O})$ in LaCoO$_3$.

\newpage

\begin{table}

{\bf TABLE I.}~Estimates of tight-binding parameters 
by a least-squared-error fitting of LMTO results 
including orbital overlaps. The numbers
in parentheses have been obtained by fitting only the transition
metal $d$ bands. \\

\begin{tabular}{l p{.5in}p{.5in}p{.5in}p{.5in}p{.5in}p{.5in}}

Compound & $sd\sigma$ (eV)&$pp\sigma$ (eV) & $pp\pi$ (eV) &
$pd\sigma$ (eV) & $pd\pi$ (eV) &
${\epsilon}_d-{\epsilon}_p$ (eV)   \\ \hline
LaTiO$_{3}$ &-2.54 & 0.58 & -0.17& -2.33 & 1.42 & 6.01 \\ 
LaVO$_{3}$  &-2.62 ~~(-2.59) & 0.73 (0.73) & -0.14 ~~(-0.14) & -2.30 ~~(-2.32)
 & 1.27 (1.28) & 4.80 (4.80) \\ 
LaCrO$_{3}$ &-2.44 ~~(-2.47) & 0.72 (0.72)& -0.16 ~~(-0.16) & -2.25
~~(-2.25) & 1.19  (1.19) & 3.86 (3.86) \\ 
LaMnO$_{3}$ &-2.21 ~~(-2.22)& 0.70 (0.70)& -0.16 ~~(-0.16)& -1.99
~~(-1.99) & 1.10  (1.10) & 3.12 (3.12) \\ 
LaFeO$_{3}$ &-1.73 ~~(-1.74) & 0.64 (0.64)& -0.15 ~~(-0.15) & -1.66
~~(-1.66) & 0.86 (0.86) & 1.98  (1.98)  \\ 
LaCoO$_{3}$ &-1.95 ~~(-1.96) & 0.77 (0.77)& -0.24 ~~(0.24)& -1.72
~~(-1.72) & 1.02 (1.02)& 1.63 (1.63) \\  
LaNiO$_{3}$ &-1.59 ~~(-1.60) & 0.76 (0.76) & -0.14 ~~(-0.14) & -1.57
~~(-1.55) & 0.97 (0.95) & 0.69 (0.75) \\ 
\end{tabular}
\end{table}

\newpage

\begin{table}

{\bf TABLE II.}~Estimates of tight-binding parameters 
by least-squared-error fitting of LMTO results with the
assumption of orthonormal basis functions. The numbers
in parentheses have been obtained by fitting only the transition
metal $d$ bands. \\

\begin{tabular}{l p{.5in}p{.5in}p{.5in}p{.5in}p{.5in}p{.5in}}

Compound & $sd\sigma$ ~~(eV) &$pp\sigma$ (eV) & $pp\pi$ (eV) &
$pd\sigma$ (eV) & $pd\pi$ (eV) &
${\epsilon}_d-{\epsilon}_p$ (eV)   \\ \hline
LaTiO$_{3}$ &-2.89& 0.52 & -0.11 & -2.50 & 1.49 & 5.91 \\ 
LaVO$_{3}$ &-2.69 ~~(-2.71)& 0.47 (0.47) & -0.07 ~~(-0.07) & -2.44
~~(-2.45) & 1.23  (1.27)& 4.70 (4.66)\\ 
LaCrO$_{3}$ &-2.34 ~~(-2.36)& 0.50 (0.50)& -0.08 ~~(-0.08)& -2.37
~~(-2.38) & 1.15 (1.18)& 3.79  (3.76)\\ 
LaMnO$_{3}$ &-2.09 ~~(-2.13)& 0.57 (0.57)& -0.10 ~~(-0.10) & -2.11
~~(-2.17) & 0.97 (1.08) & 3.10 (2.93)\\ 
LaFeO$_{3}$ &-1.68 ~~(-1.67)& 0.50 (0.50) & -0.08 ~~(-0.08) & -1.72
~~(-1.72) & 0.82 (0.83) & 1.85  (1.85) \\ 
LaCoO$_{3}$ &-1.69 ~~(-1.72) & 0.62 (0.62) & -0.12 ~~(-0.12) & -1.91
~~(-1.92) & 0.89 (0.92) & 1.50  (1.44) \\  
LaNiO$_{3}$ &-1.40 ~~(-1.62) & 0.66 (0.66) & -0.12 ~~(-0.12) & -1.67
~~(-1.73) & 0.77 (0.88) & 0.69 (0.4) \\ 
\end{tabular}
\end{table}
\end{document}